# 含风光出力随机性的独立微电网二次频率控制

钟诚[1,2]，姜志富[2]，张翔宇[2]，陈继开[1,2]，李扬[2]

(1.现代电力系统仿真控制与绿色电能新技术教育部重点实验室（东北电力大学）吉林省 吉林市 132012;

2.东北电力大学电气工程学院，吉林省吉林市，132012)

**摘　要：**微电网包含风、光发电等分布式电源，具有惯性小、随机性大等特点，给系统频率控制带来挑战。然而，如果风、光伏发电采用有功备用控制，可预留部分功率参与频率调节，提高系统频率控制能力。针对风、光发电随机性的问题，提出一种考虑新能源减载参与二次调频的模型预测控制方法。建立包含随机功率扰动的扩展状态矩阵，采用卡尔曼滤波估算随机未知扰动；依据风、光最大可用功率，建立实时变约束，避免机组功率越限；设置合理的权重系数，优先风、光发电出力参与二次调频；通过求解变约束二次规划问题，获得各个机组的优化调频功率。最后，建立含多个光伏、风电的微电网模型，在不同场景下与常规二次调频方法进行对比仿真。仿真结果表明，所提方法能够提高系统频率恢复速度，减小系统频率波动，尤其在风、光发电剧烈波动场景下。

## 0 引 言

随着国家“双碳”战略目标制定，新能源发电得到进一步重视，将成为未来主要发电形式[1]。微电网能够有效的整合各种分布式发电和储能，实现负荷就近供电，为提高新能源接入水平提供一种途径，是未来电网的重要组成部分[2]。

微电网可以工作在并网和孤岛两种模式[3]。孤岛模式下，微电网频率稳定需要依靠其包含的分布式发电机组调节[4]。但这些电源惯性小，出力随机性大（例如风力、光伏发电）。因此，微电网需求更加先进灵活控制方法，来提高频率控制能力[5]。类似于常规大电网，微电网频率调节可以分成一次调频，二次调频，甚至三次调频[6-7]。一次调频为本地有差调节，无法实现频率恢复。引入二次调频控制，可进一步调整分布式发电出力，提高频率品质。

常规二次调频通过采用 PI 控制器帮助系统频率恢复。但是微电网中电源种类差异较大，集中 PI 控制不能灵活整合各种调频资源。文献[8]采用 PI 控制器来实现二次调频控制，但是其参数整定受限于非机理模型的准确性。文献[9]提出了一种基于 ADP（Adaptive dynamic programming）的控制策略，该方法考虑了分布式电源出力的不确定性以及负荷的随机性，调整柴油机和储能系统的出力来抑制频率波动。文献[10]提出了能够将一、二次调频进行切换的 VSG 控制，但没有考虑变流器响应速度不同对频率调整的影响。文献[11]提出了一种多微电网系统鲁棒模型预测控制方法。该方法利用系统之间的耦合来减小微电网系统的频率波动，从而提高系统的稳定性。文献[12]提出了一种自适应模型预测控制方法，用于确保有新能源发电装置的两区域互联的频率控制。

近年来，模型预测控制（Model Predictive Control，MPC）得到迅速发展，其滚动优化特点可以较好的处理风光实时变化的出力所引起的频率波动。文献[13]提出了一种模糊控制结合 MPC 来控制 VSG 的控制方式，模糊 MPC 控制器通过修正虚拟惯量和阻尼系数，从而提高孤岛微电网的频率稳定性。但是模糊控制主要依赖于经验和凑试，并且控制规则一旦确定，不能实时调整，不太适合处理随机性较强的环境。文献[14]提出了一种可以优化微电网中的储能电池之间的功率潮流的模型预测控制策略，同时使用求解器进行快速处理储能电池的非线性变化，可以提高实时 MPC 的运算速度，保证控制器更加快速的调节频率。但是该文献仅仅考虑了储能系统，若系统新能源渗透率较高、波动较大时，只有储能参与调频可能会出现频率越限问题。文献[15]针对于 VSG 二次调频能力不足的问题提出了 MPC 的控制方式，来减少频率恢复时间。但是，该论文研究的也仅仅是结合 VSG 和 MPC 的控制方式，针

国家自然科学基金(52077030), 吉林省自然科学基金 (20190201289JC, YDZJ202101ZYTS149)

National Natural Science Foundation of China (52077030); Natural Science Foundation of Jilin Province (20190201289JC, YDZJ202101ZYTS149)

对于微电网的新能源随机性引起的频率问题没有深入的讨论，上述研究多集中于传统储能装置，没有涉及到新能源发电机组参与调频情况。近年来，风、光发电有功备用参与系统频率得到关注[16-18]。风、光发电采用减载模式预留部分有功功率，参与系统频率调节改善频率响应。但风、光发电具有随机性，使其参与二次调频增加难度。

综上，提出了一种考虑新能源减载参与二次调频控制方法。该方法采用模型预测控制（MPC）构架，实时优化各个分布式单元出力。为缓解风、光发电随机的影响，建立包含随机扰动的扩展状态矩阵，采用卡尔曼滤波估算随机扰动；依据实时估算风、光最大功率估算值，建立实时变约束。另外，通过合理权重设计，优先风、光发电出力。最后，建立微电网仿真模型，在不同场景下与传统控制策略进行对比验证。

# 1 微电网模型

## 1.1 微电网结构

独立微电网结构如图 1 所示。主要包含两个分布式光伏发电机组（PV1，PV2），两个分布式风力发电机组(WT1 ,WT2)，储能单元（BESS）和柴油机组（DU），及其对应的负荷。各发电单元的容量配置如下表 1 所示

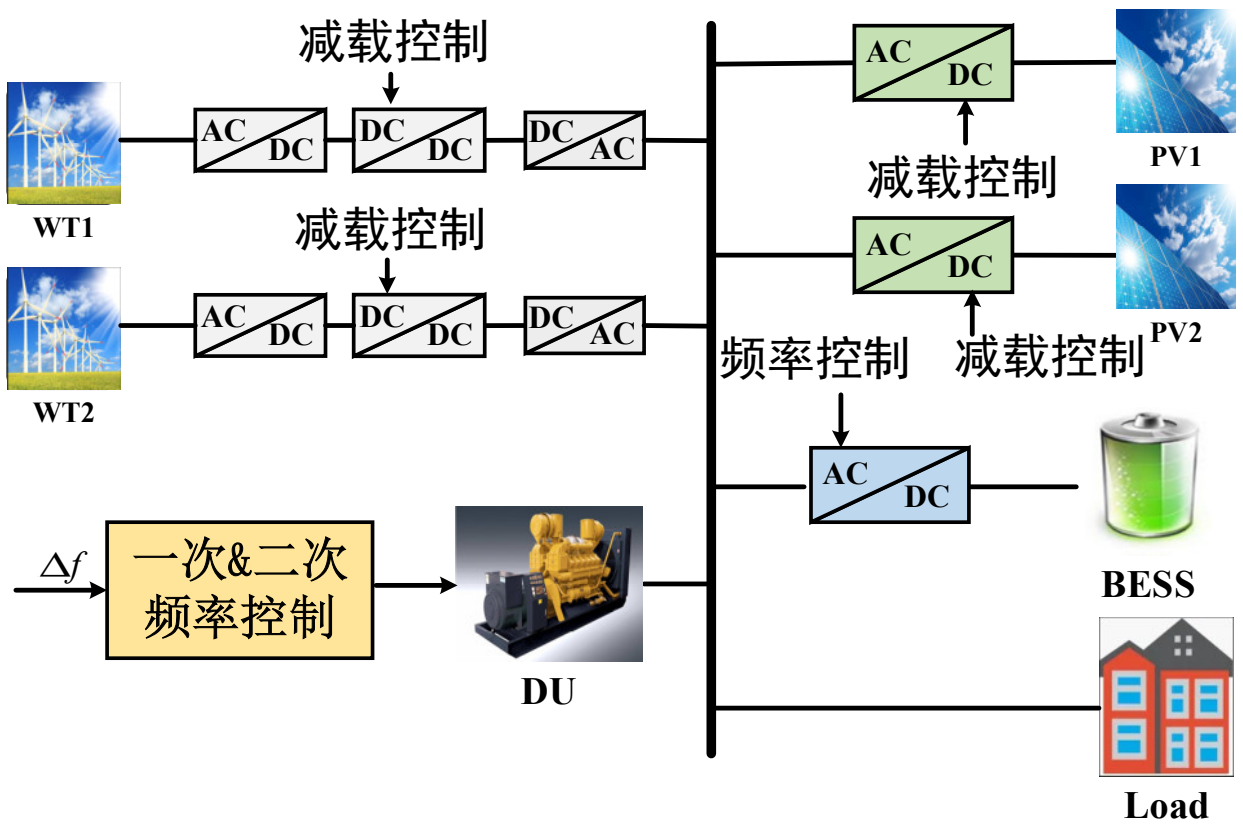

图 1 微电网结构示意图

Fig.1 Schematic diagram of microgrid structure

微电网独立运行时，需要依靠自身发电单元调节，维持电压和频率稳定。微电网频率控制包含一次调频和二次调频两层。一次调频为本地控制，发电单元通过下垂控制，依据频率波动改变机组有功出力，减少功率扰动下的频率偏差。但一次调频为有差调节，为维持微电网频率恒定，需要增加二次调频控制。二次调频控制协调多个电源出力，进一步消除频率偏差。常规二次调频采用低带宽通信，速度较慢集中 PI 控制器实现。

通常，独立微电网频率调节由储能和柴油机承担。而光伏、风电等单元受发电随机性影响，常采用最大功率控制，不参与微电网频率调节。近年来，风、光主动参与系统频率调节的研究逐渐兴起。风、光发电采用减载控制，预留部分有功功率。当系统频率变化时，可以灵活调节该部分预留功率，改善系统频率响应。但是，目前风、光减载调频策略主要集中在一次调频策略。受风、光发电随机性的影响，其备用功率处在实时变化中，给风、光参与二次调频增加难度。

**表 1 微电网配置**

Tab.1 Microgrid configuration

| 微电网电源 | 符号 | 数值 | 单位 |
|---|---|---|---|
| 光伏阵列 | $P_{PV1}$ | 80 | kW |
| | $P_{PV2}$ | 80 | kW |
| 风机 | $P_{wt1}$ | 60 | kW |
| | $P_{wt2}$ | 60 | kW |
| 储能系统 | $P_{BESS}$ | 100 | kW |
| 柴油机 | $P_{DU}$ | 120 | kW |
| 负荷 | $P_{Load}$ | 200 | kW |

## 1.2 微电网状态空间模型

主要聚焦考虑风、光减载下的微电网二次调频控制。结合图 1，微电网的负载频率控制（LFC）模型如下图 2 所示。

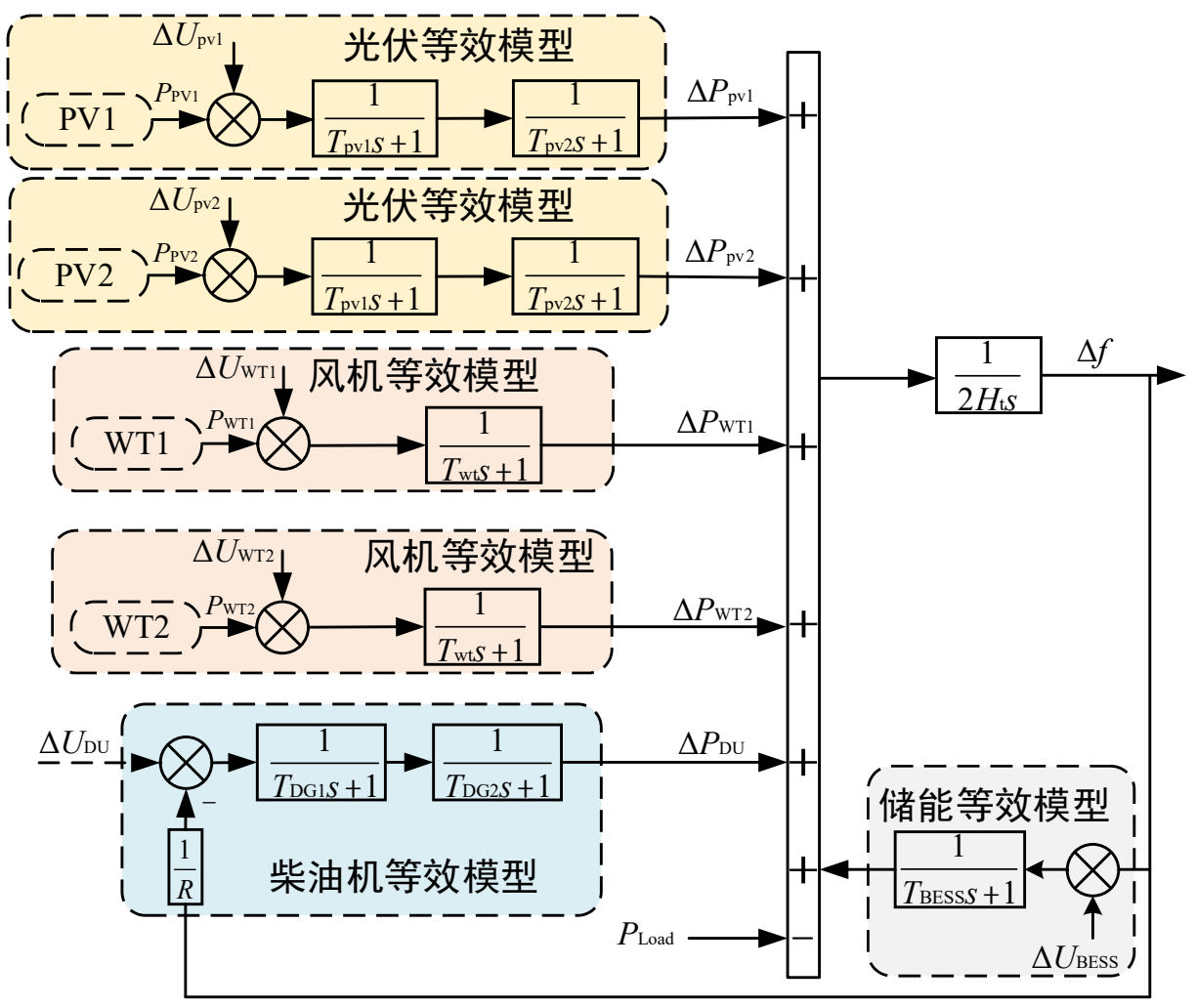

图 2 微电网等效模型

Fig.2 Microgrid equivalent model

微电网中，分布式光伏、风电和储能单元采用并网逆变器接口接入电网。通过并网逆变器控制，DG 单元可以快速跟踪给定参考功率。在 LFC 模型中，为了简化，采用一阶模型表示并网

逆变器模型。关于更多 LFC 模型的细节，可以参考论文[19]。

图 2 中，$T_{pv1}$ 为光伏阵列时间常数，$T_{pv2}$ 为光伏逆变器控制时间常数，$\Delta U_{pv}$ 为光伏控制输入参考功率；$T_{wt}$ 为风机时间常数，$\Delta U_{WT}$ 为风机控制输入；$R$ 和 $T_{DU1}$ 分别为柴油机下垂系数和调速时间常数，$T_{DU2}$ 为柴油机的时间常数。$T_{BESS}$ 为储能系统的时间常数，$R$ 为柴油机下垂系数。上述参数具体值如下表 2 所示。

**表 2 微电网参数**

Tab.2 Microgrid parameters

| 微电网电源 | 符号 | 数值 | 单位 |
|---|---|---|---|
| 光伏阵列 | $T_{pv1}$ | 0.15 | s |
| | $T_{pv2}$ | 0.08 | s |
| 风机 | $T_{wt}$ | 0.3 | s |
| 储能系统 | $T_{BESS}$ | 0.1 | s |
| | $T_{DU1}$ | 0.4 | s |
| 柴油机 | $T_{DU2}$ | 0.1 | s |
| | $R$ | 3 | Hz/pu(kW) |
| 微电网惯性 | $H_t$ | 0.6 | s |

进一步，将图 2 微电网 LFC 模型，整理为状态空间形式，如下式(1)所示。

$$\begin{cases}\dot{x}(t)=A_c x(t)+B_{cu}u(t)+D_{ca}d(t)\\ y_c(t)=C_c x(t)\end{cases} \tag{1}$$

式中：$x$ 为状态变量，$u$ 为控制变量，$d$ 为风、光出力波动以及负荷扰动，$y$ 为输出变量; $A_c$，$B_c$，$C_c$，$D_c$ 分别为连续状态方程的状态常数矩阵，控制常数矩阵，扰动常数矩阵和输出常数矩阵。式(1)具体展开式见附录 1 所示。

## 2 随机性的观测器设计

### 2.1 随机性观测器结构

风光的不确定性以及负荷扰动对于 MPC 控制器来说是未知扰动 $d(k)$，会直接影响 MPC 的控制效果，所以首先要对未知扰动进行观测处理，将系统的状态空间表达式进行解耦，分解为两部分，一部分为系统中的已知状态变量 $\boldsymbol{x}_1$ 和未知状态变量 $\boldsymbol{x}_2$，其 $\boldsymbol{x}_2$ 表示为受未知扰动影响的频率偏差 $\Delta f(\mathrm{k})$。

将式（1）进行离散化处理，可得：

$$\begin{cases}x(k+1)=Ax(k)+Bu(k)+Dd(k)\\ y(k)=Cx(k)\end{cases} \tag{2}$$

式(2)中，$A$，$B$，$C$，$D$ 分别是离散状态方程的状态矩阵、控制矩阵，扰动矩阵和输出矩阵。上述系数矩阵通过连续状态方程（1）中的系数矩阵离散化获得。采用 DU Hamel 方法进行离散化[20]，如下式。

$$A=\mathrm{e}^{A_c T_s} \tag{2a}$$

$$B=\int_0^{T_s}\mathrm{e}^{A_c\tau}\mathrm{d}\tau\cdot B_{cu} \tag{2b}$$

$$D=\int_0^{T_s}\mathrm{e}^{A_c\tau}\mathrm{d}\tau\cdot D_{cd} \tag{2c}$$

式中 $T_s$=0.2s 为系统采样时间。

对于解耦的状态变量，定义其中 $N$ 为任意矩阵，使得$\psi=[N\ D_{cd}]$是非奇异矩阵，将$\psi$与离散化之后的（2）相乘，得：

$$\begin{cases}\bar{x}(k+1)=\bar{A}\bar{x}(k)+\bar{B}u(k)+\bar{D}d(k)\\ y(k)=\bar{C}\bar{x}(k)\end{cases} \tag{3}$$

$$\bar{A}=\begin{bmatrix}\bar{A}_{11} & \bar{A}_{12}\\ \bar{A}_{21} & \bar{A}_{22}\end{bmatrix}=\Psi^{-1}A\Psi\text{；}\ \bar{B}=\begin{bmatrix}\bar{B}_1 & \bar{B}_2\end{bmatrix}^{\mathrm{T}}=\Psi^{-1}B\text{；}$$

$$\bar{D}=\Psi^{-1}D\text{；}\quad \bar{C}=C\Psi=\begin{bmatrix}CN & CD\end{bmatrix}\text{；}$$

$$\bar{x}=\begin{bmatrix}\bar{x}_1 & \bar{x}_2\end{bmatrix}^{\mathrm{T}}=\Psi^{-1}x=\Psi^{-1}\begin{bmatrix}x_1 & x_2\end{bmatrix}^{\mathrm{T}}\text{；}$$

因此，可以得到对应于已知得状态变量 $\boldsymbol{x}_1$ 的表达式：

$$\begin{cases}\begin{bmatrix}1 & 0\end{bmatrix}^{\mathrm{T}}\bar{x}(k+1)=\begin{bmatrix}\bar{A}_{11} & \bar{A}_{12}\end{bmatrix}\bar{x}(k)+\bar{B}_1u(k)\\ y(k)=\begin{bmatrix}CN & CD\end{bmatrix}\bar{x}(k)\end{cases} \tag{4}$$

状态变量 $\boldsymbol{x}_2$ 可以从测量输出 $y(k)$中获得，式（4）可以表示为一个线性表达式。

在转移矩阵 $U=\begin{bmatrix}CD & \Gamma\end{bmatrix}$中，$CD$ 是一个全列秩矩阵，$\Gamma$ 是一个任意矩阵，因此 $U$ 是一个非奇异矩阵。因此，$U^{-1}=[U_1\ U_2]^{\mathrm{T}}$，将式（3）中的测量方程乘以 $U^{-1}$，可以得到：

$$U_1y(k)=U_1CN\bar{x}_1(k)+\bar{x}_2(k) \tag{5}$$

$$U_2y(k)=U_2CN\bar{x}_1(k) \tag{6}$$

将公式（5）代入公式（4）并与其（6）结合，可得公式（7）：

$$\begin{cases}\bar{x}_1(k+1)=\tilde{A}\bar{x}_1(k)+\bar{B}_1u(k)+Ey(k)\\ \bar{y}(k)=\tilde{C}\bar{x}_1(k)\end{cases} \tag{7}$$

式中：$\tilde{A}=\bar{A}_{11}-\bar{A}_{12}U_1CN$ 为修正状态矩阵，$\tilde{C}=U_2CN$ 为修正测量矩阵，$\bar{y}(k)=U_2y(k)$ 为修正测量向量，$E=\bar{A}_{12}U_1$。

如果 $\tilde{A}$、$\tilde{C}$ 是可观测的，则可以设计卡尔曼

滤波器。系统可观测性的存在条件在文献[26]中进行了检验。卡尔曼滤波器如（8）所示。

$$\hat{\bar{x}}_1(k+1|\, k)=(\tilde{A}-L\tilde{C})\hat{\bar{x}}_1(k|\, k-1)+\bar{B}u(k)+L^{*}\bar{y}(k) \quad （8）$$

其中：$L^{*}=LU_2+E$，$L$ 是卡尔曼增益矩阵

利用式（5）和（8）就可以估计出系统中所有的状态变量，即

$$\hat{x}(k|\, k)=\Psi\hat{\bar{x}}=\Psi\begin{bmatrix}\hat{\bar{x}}_1(k|\, k)\\ \hat{\bar{x}}_2(k|\, k)\end{bmatrix} \quad （9）$$

其中 $\hat{\bar{x}}_2(k|\, k)=U_1 y(k)-U_1 CN\hat{\bar{x}}_1(k|\, k)$。表示 $\Delta f$ 可以从测量值和剩余的估计状态 $\boldsymbol{x}_1$(k)和测量值 $y(k)$中估计出来[27]。

为了识别未知扰动 $d$(k)，扰动包括风机出力、光伏出力和负荷扰动。将（9）代入到（3）中的第一个表达式，可得

$$\hat{d}(k)=F_1\bar{y}(k+1)+F_2\hat{\bar{x}}_1(k|\, k)+F_3 y(k)+F_4 u(k) \quad （10）$$

式中：

$$\begin{cases}F_1=U_1\\ F_2=U_1CNLU_2CN+U_1CN\bar{A}_{12}U_1CN\\ \quad -U_1CN\bar{A}_{11}-\bar{A}_{21}+\bar{A}_{22}U_1CN\\ F_3=-U_1CNLU_2-U_1CN\bar{A}_{12}U_1-\bar{A}_{22}U_1\\ F_4=-U_1CN\bar{B}_1-\bar{B}_2\end{cases}$$

## 3 考虑随机性的模型预测控制器的设计

### 3.1 MPC 控制器结构

微电网二次调频控制为多输入单输出模型，系统存在强耦合。且风机、光伏出力及负荷的不确定性波动会引起系统频率波动。常规 PI 控制并不适合该类系统的控制。

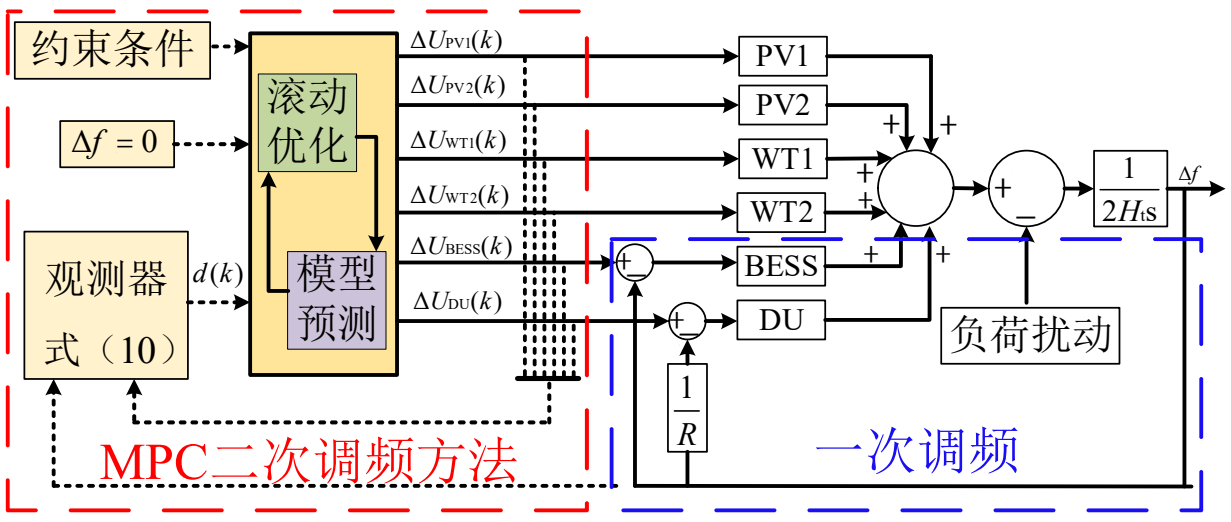

图 3 MPC 控制的微电网频率调整结构示意图

Fig.3 Schematic diagram of MPC controlled microgrid frequency

模型预测控制（MPC）在处理多输入多输出耦合系统时，可获得令人满意的控制效果。采用 MPC 控制来实现微电网二次调频控制，其控制结构如图 3 所示。将系统输出和控制器控制量反馈到观测器，通过式（10）进行不确定性观测。首先将估计出来的 $d(k)$ 输入到模型预测环节中，其次在考虑到约束条件和频率波动参考值 $\Delta f$，设置优化目标进行求解，最后对优化目标进行滚动式优化，进一步输出控制量。

图 3 中，MPC 控制器频率波动参考值为 0，输出为各个分布式电源的调节功率输入参考值 $\Delta U$。采用卡尔曼滤波器获得系统当前状态量 $\boldsymbol{x}$(k|k)。模型预测控制依据当前状态量和系统离散模型，预测有限时域内系统状态量；采用滚动最优控制的思想，求解满足约束下最优控制量，输出 $\Delta U$ 调整电源单元出力，改善系统频率响应。

### 3.2 MPC 控制器设计

（1）模型预测

考虑到 MPC 的准确性以及计算的复杂性，设定预测时域 $p$=10，控制时域 $m$=3。结合系统离散模型（2），可得到预测时域 $p$ 内的频率偏差为：

$$Y_{\rm p}(k+1|\, k)=S\Delta x(k)+Iy(k)+S_{\rm B}\Delta U(k) \quad （11）$$

$$S^{\rm T}=\begin{bmatrix}S_{\rm x}^{\rm T}\\ S_{\rm d}^{\rm T}\end{bmatrix} \quad （11a）$$

$$S_{\rm x}=\begin{bmatrix}CA & \sum_{i=1}^{2}CA^{i} & \sum_{i=1}^{3}CA^{i} & \cdots & \sum_{i=1}^{p}CA^{i}\end{bmatrix}^{\rm T} \quad （11b）$$

$$I=\begin{bmatrix}1 & 1 & \cdots & 1 & 1\end{bmatrix}_{1\times {\rm p}}^{\rm T} \quad （11c）$$

$$S_{\rm B}=\begin{bmatrix}CB & 0 & 0\\ \sum_{i=1}^{2}CA^{i-1}B & CB & 0\\ \sum_{i=1}^{3}CA^{i-1}B & \sum_{i=1}^{2}CA^{i-1}B & CB\\ \sum_{i=1}^{4}CA^{i-1}B & \sum_{i=1}^{3}CA^{i-1}B & \sum_{i=1}^{k-m+1}CA^{i-1}B\\ \vdots & \vdots & \vdots\\ \sum_{i=1}^{p}CA^{i-1}B & \sum_{i=1}^{p-1}CA^{i-1}B & \sum_{i=1}^{p-m+1}CA^{i-1}B\end{bmatrix} \quad （11d）$$

$$S_{\rm d}=\begin{bmatrix}CD & 0 & 0\\ \sum_{i=1}^{2}CA^{i-1}D & CD & 0\\ \sum_{i=1}^{3}CA^{i-1}D & \sum_{i=1}^{2}CA^{i-1}D & CB\\ \sum_{i=1}^{4}CA^{i-1}D & \sum_{i=1}^{3}CA^{i-1}D & \sum_{i=1}^{k-m+1}CA^{i-1}D\\ \vdots & \vdots & \vdots\\ \sum_{i=1}^{p}CA^{i-1}D & \sum_{i=1}^{p-1}CA^{i-1}D & \sum_{i=1}^{p-m+1}CA^{i-1}D\end{bmatrix} \quad （11e）$$

式中：$Y_{\rm p}$（$k$+1|$k$）为预测时域内的第 k 时刻频率偏差，$k$=1，2，3…P。

（2）约束条件

对于柴油机和储能，其调频可用功率受其功

率容量的约束。因此，

$$\begin{cases} P_{\text{DU,min}} - \bar{P}_{\text{DU}} \le \Delta U_{\text{DU}}(k) \le P_{\text{DU, max}} - \bar{P}_{\text{DU}} \\ P_{\text{BESS,min}} - \bar{P}_{\text{BESS}} \le \Delta U_{\text{BESS}}(k) \le P_{\text{BESS,max}} - \bar{P}_{\text{BESS}} \end{cases} \quad (12)$$

式中，$P_{\text{DU,min}}$、$P_{\text{DU, max}}$、$P_{\text{BESS,min}}$ 和 $P_{\text{BESS,max}}$ 分别为柴油机或储能单元的功率上、下限值。$\bar{P}_{\text{DU}}$，$\bar{P}_{\text{BESS}}$ 为当前时刻上层调度功率值。

风、光发电可用功率受外部环境和减载水平 $d$%影响。选取风电和光伏减载水平 $d$%为 10%。

风电机组发出功率可简化表示为[21]，式中，

$$\begin{cases} P_{\text{wt}} = \dfrac{1}{2}\rho_{\text{air}}\pi R^2 v_{\text{w}}^3 C_{\text{p}} \\ C_{\text{p}} = c_1\left(c_2 z - c_3\beta - c_4\right)e^{c_5 z} \\ z = \dfrac{1}{\lambda + c_6\beta} - \dfrac{c_7}{1+\beta^3} \\ \lambda = \dfrac{\omega_{\text{M}} R}{v_{\text{w}}} \end{cases} \quad (13)$$

$P_{\text{wt}}$ 为双馈风机功率输出；$\rho$ 为空气密度；R 为双馈叶片半径；$v$ 是风速；$C_{\text{p}}$ 为双馈桨叶功率系数，其值与桨距角 $\beta$ 和叶尖速比 $\lambda$ 有关，当 $\beta$ 不变时，可以通过改变 $\lambda$ 来调整 $C_{\text{p}}$ 的大小，来获得风能利用系数最优值 $C_{\text{p-opt}}$。$C_{\text{p-opt}}$ 是随着桨距角 $\beta$ 减小而增大，通过贝兹理论可以知道，风能利用系数的极限值为 16/27；$C_1$−$C_7$ 是风机特性的相关参数；$\omega$ 为叶片角速度，其值与转子转速 $\omega_{\text{r}}$ 成正比。

其风机的发出功率 $P_{\text{wt}}$ 可表示为：

$$P_{\text{MAP,WT}} = \begin{cases} 0 & \omega \le \omega_{\min} \\ (1-d\%)P_{\text{wt}} & \omega_{\min} < \omega \le \omega_{\max} \\ (1-d\%)P_{\text{wt}} & \omega \ge \omega_{\max} \end{cases} \quad (14)$$

光伏与风机的控制方式类似，也采用最大功率跟踪控制策略，其最大功率估算表达式为[21]：

$$\begin{cases} P_{\text{PV}} = \left(n_{\text{s}} n_{\text{p}}\right)\cdot P_{\text{g}}^* \dfrac{G_{\text{eff}}}{G^*}\left(1-\gamma\left(T_{\text{c}} - T_{\text{c}}^*\right)\right) \\ T_{\text{c}} = T_{\text{a}} + C_{\text{T}} G_{\text{eff}} \\ C_{\text{T}} = \dfrac{\text{NOCT} - 20}{0.8G^*} \end{cases} \quad (15)$$

式中，$P_{\text{MPPT,PV}}$ 为光伏阵列最大输出功率；$n_{\text{s}}$，$n_{\text{p}}$ 分别为光伏组件的串联和并联数；$P_{\text{g}}^*$ 为光伏组件的额定峰值功率；$G_{\text{eff}}$ 为有效太阳辐照度；$G^*$ 为标准太阳辐照度，取值为 1000W/m$^2$；$\gamma$ 为功率温度系数；$T_{\text{c}}$ 为光伏组件的工作温度；$T_{\text{c}}^*$ 为标准测量温度，取值为 25℃；$T_{\text{a}}$ 为环境温度；$C_{\text{T}}$ 为温度辐射系数；NOCT 为光伏组件的额定温度。

光伏机组的发出功率 $P_{\text{pv}}$ 可表示为：

$$P_{\text{MAP,PV}} = \left(1 - d_{\text{pv}}\%\right)P_{\text{PV}}\left(G_{\text{eff}}, T_{\text{a}}\right) \quad (16)$$

由上述式(11)和(13)可知。获得风、光发电机组的调频可用功率风电关键是获得机组最大可用功率 $P_{\text{MPPT}}$。

具体的风力、光伏发电机组减载运行控制，已经有较多文献讨论[22-25]，不再赘述。据此，风、光发电机组可用调频功率约束设置为：

$$\begin{cases} -d\%P_{\text{MAP,WT}} \le \Delta U_{\text{WT}}(k) \le d\%P_{\text{MAP,WT}} \\ -d\%P_{\text{MAP,PV}} \le \Delta U_{\text{PV}}(k) \le d\%P_{\text{MAP,PV}} \end{cases} \quad (17)$$

（3）优化目标

当频率出现波动时，MPC 需要对系统频率偏差 $Y_{\text{p,c}}$ 和输出控制量 $\Delta U$ 进行权衡，即存在频率误差时，控制器发出控制指令 $\Delta U$ 使得频率恢复到额定值，具体表达式为：

$$J(x(k), \Delta U(k)) = \left\|\boldsymbol{\Gamma}_{\text{y}} Y_{\text{p,c}}(k)\right\|^2 + \left\|\boldsymbol{\Gamma}_{\text{u}}\Delta U(k)\right\|^2 \quad (18)$$

式中：$\boldsymbol{\Gamma}_{\text{y}}$ 和 $\boldsymbol{\Gamma}_{\text{u}}$ 分别为频率偏差和控制器输出变的权重矩阵。$Y_{\text{p,c}}(k)$ 表示预测时域内第 k 时刻频率偏差，$\Delta U(k)$ 为 MPC 第 k 时刻控制器输出，即系统调频功率。

$$\begin{cases} \Gamma_{\text{y}} = diag\{\alpha \quad \alpha \quad \alpha \quad \cdots \quad \alpha\}_{10\times 10} \\ \Gamma_{\text{u}} = diag\{\beta_{\text{PV1}} \quad \beta_{\text{PV2}} \quad \beta_{\text{WT1}} \quad \beta_{\text{WT2}} \quad \beta_{\text{DU}} \quad \beta_{\text{BESS}}\} \end{cases} \quad (19)$$

式（19）中，$\alpha$ 和 $\beta$ 为对应的惩罚因子。$\boldsymbol{\Gamma}_{\text{y}}$ 用来惩罚系统频率偏差，$\boldsymbol{\Gamma}_{\text{u}}$ 用来惩罚控制输出，即发电单元输出功率。$\alpha$=1.6596；$\beta_{\text{WT}}$=$\beta_{\text{PV}}$ =0.2894，风、光取相同权重。$\beta_{\text{DU}}$=$\beta_{\text{BESS}}$=0.3762，柴、储单元比风、光发电单元权重值大，这样处理可以优先风、光机组参与二次调频。

（4）含约束最优化问题求解

由于约束条件的存在，不能直接得到目标函数的最优解。因此，需要将变约束的 MPC 优化问题转换成二次规划问题。将预测方程式（8）带入目标函数中，由于 MPC 只有控制变量 $U$ 有关，所以对于优化问题而言，目标函数简化为：

$$\begin{aligned} \tilde{J} = {} & \Delta U(k)^{\text{T}}(S_{\text{u}}^{\text{T}}\boldsymbol{\Gamma}_{\text{y}}^{\text{T}}\boldsymbol{\Gamma}_{\text{y}} S_{\text{u}} + \boldsymbol{\Gamma}_{\text{u}}^{\text{T}}\boldsymbol{\Gamma}_{\text{u}})\Delta U(k) \\ & -(2S_{\text{u}}^{\text{T}}\boldsymbol{\Gamma}_{\text{y}}^{\text{T}}\boldsymbol{\Gamma}_{\text{y}} E_{\text{p}}(k+1|\,k))^{\text{T}}\Delta U(k) \end{aligned} \quad (20)$$

将约束条件转换为不等式形式：

$$C_u\Delta U(k)\ge b(k+1|k) \tag{21}$$

为了求解二次规划形式的最优问题，定义新的变量，即：

$$\rho=\begin{bmatrix}\Gamma_y\left(Y_p(k+1)-R(k+1)\right)\\ \Gamma_u\Delta U(k)\end{bmatrix} \tag{22}$$

优化问题就变成了 $min\ \rho^T\rho$，由 $\rho^T\rho=(Az-b)^T(Az-b)$的极值条件，求导可得极值解为：

$$Z^*=(A^TA)^{-1}A^Tb \tag{23}$$

可知式（19）取得最小值的解，即 k 时刻控制序列最优解为：

$$\Delta U^*(k)=K_{mpc}(0-Y_p(k+1|k)) \tag{24}$$

$$K_{mpc}=\left(\mathcal{S}_u^T\Gamma_y^T\Gamma_y\mathcal{S}_u+\Gamma_u^T\Gamma_u\right)^{-1}\mathcal{S}_u^T\Gamma_y^T\Gamma_y \tag{25}$$

只取最优控制序列中第一个元素作为输出。

（5）MPC 控制器流程

综上所述，MPC 控制器流程图如下所示。

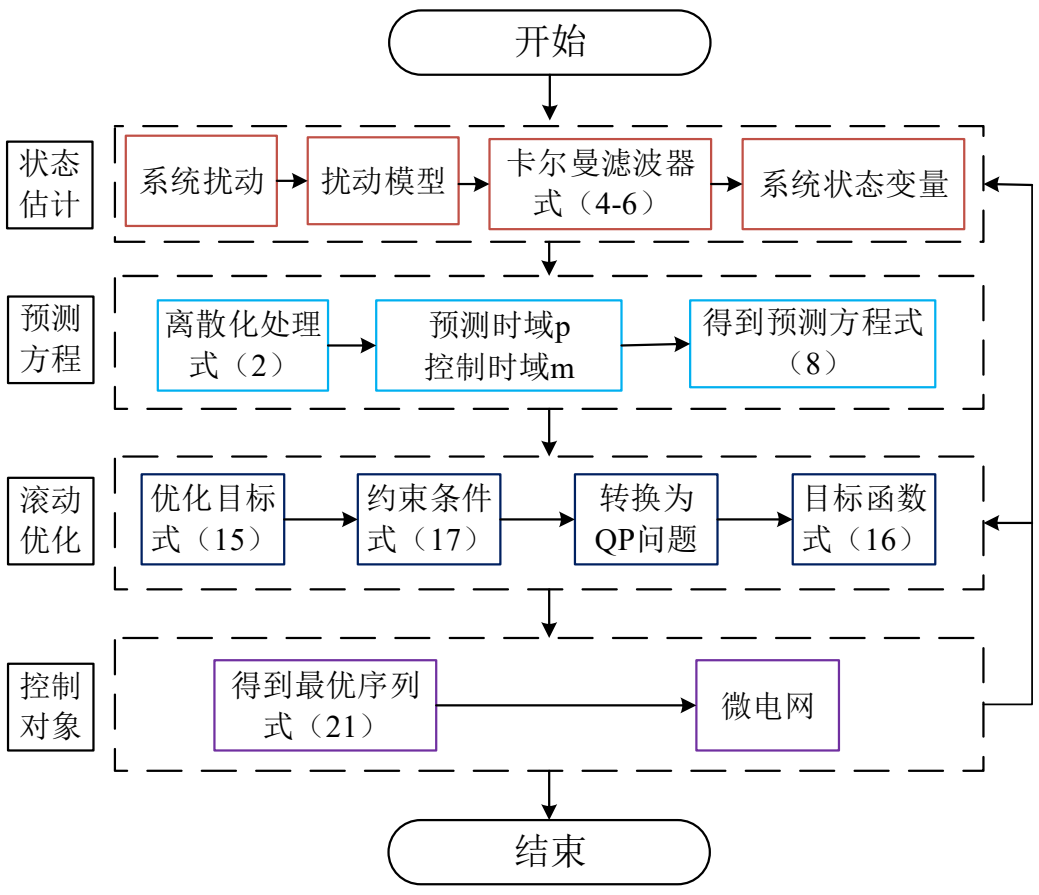

图 4 约束 MPC 实现流程图

Fig.4 Constrained MPC implementation flow chart

## 4 算例分析

为验证所提出控制策略的正确性，依照图 1，建立孤岛型风光柴储微电网仿真模型。控制方法分别采用 PI 控制（只有柴、储参与）、PI 控制（风、光、柴、储共同参与）和所提 MPC 控制三种方式。

### 4.1 风速、光照不变，负荷突变

假定风速，光照恒定，负荷阶跃突变仿真结果见图 5a~图 5f，其中实线代表实际负荷波动，虚线代表对于扰动的观测辨识。

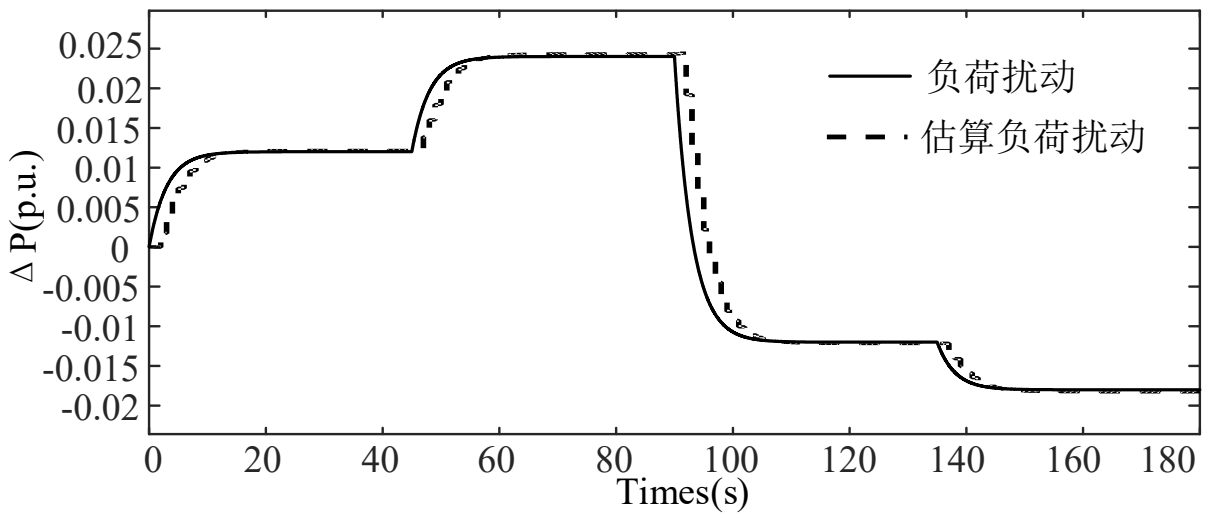

a.负荷波动

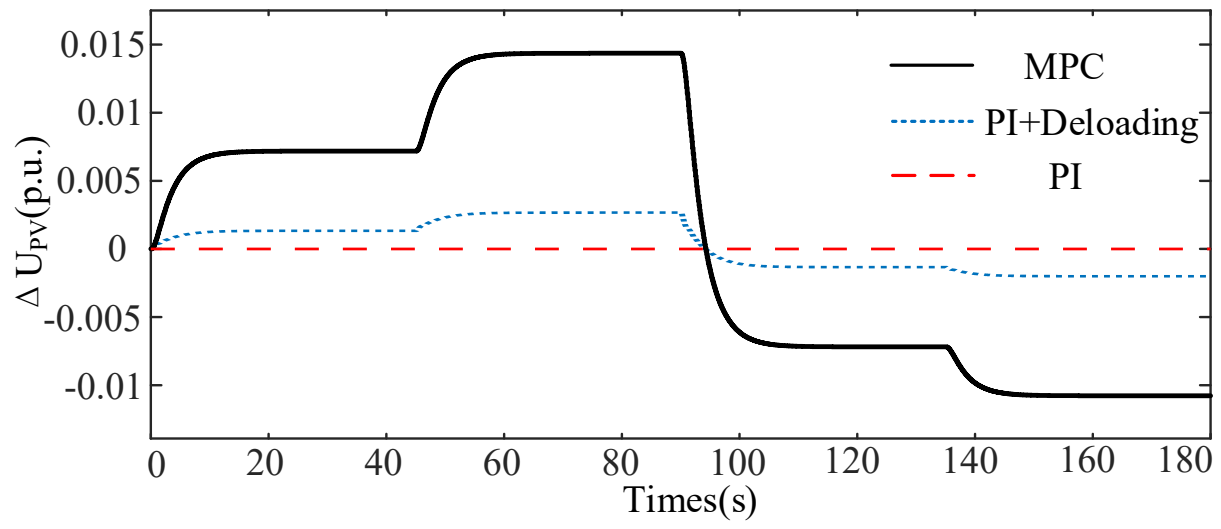

b.光伏出力

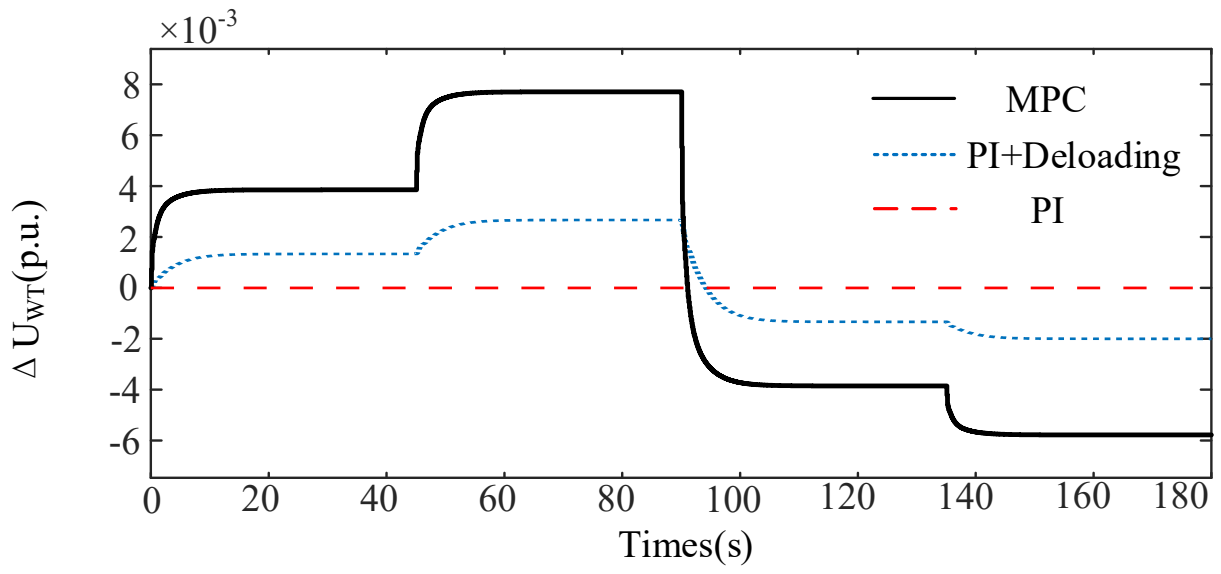

c.风机出力

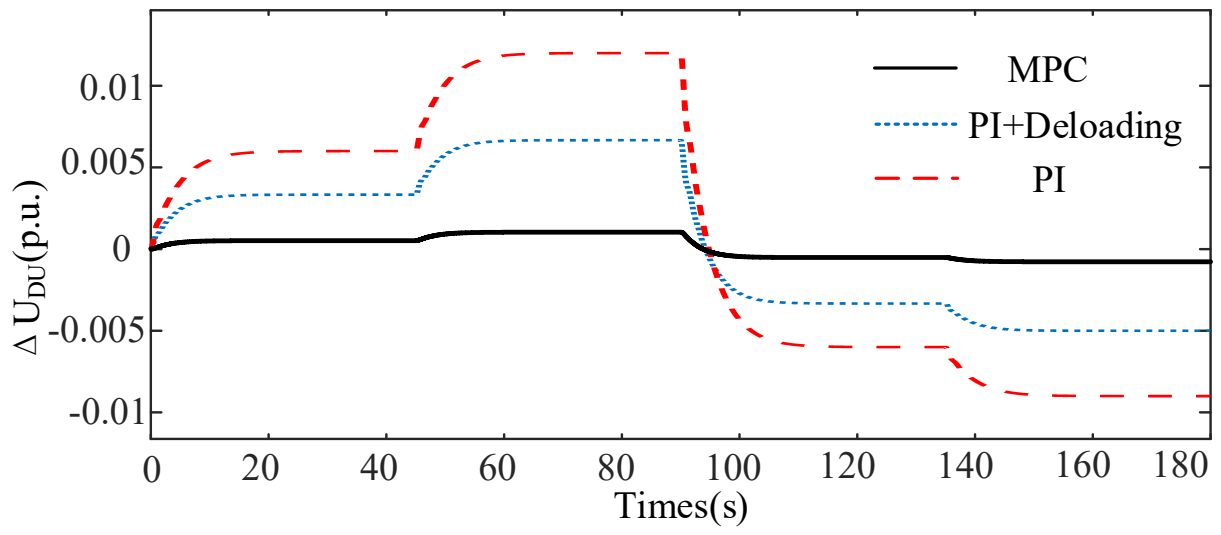

d.柴油机出力

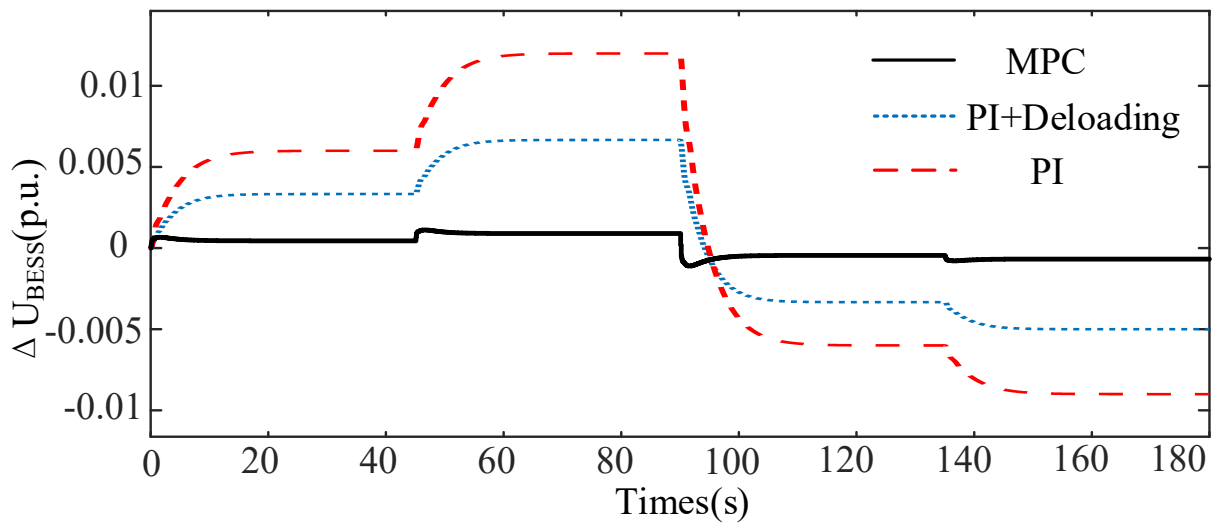

e.储能系统出力

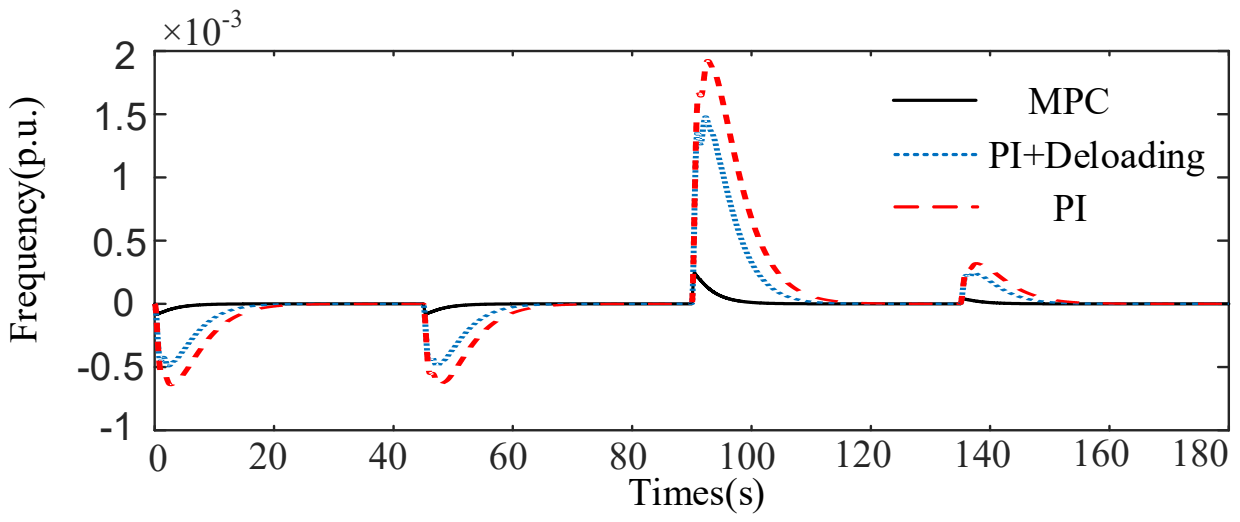

f.系统频率

图 5 案例 1 结果

Fig.5 Case 1 Results

图 5b~图 5e 中，两种 PI 控制的二次调频出力按机组容量分配，各个发电单元出力波形形状一致。相比于 PI 二次调频控制，减载+PI 控制中，新能源机组参与调频，改善了频率效果。进一步，所提 MPC 控制方法并结合约束和权重，通过滚动优化协调各个发电单元出力。由于光伏和风机出力惩罚因子小于柴油机和储能。因此光伏和风机调频出力大于减载+PI 控制方法，柴油机和储能调频出力相对应小，优先新能源机组调频。

图 5a 中，虚线代表为预测的负荷波动，当负荷变化时，由于预测需要系统输出值，所以导致预测会有一定的时滞性，但是最终会与实际值一致。图 5f 中，相对于两种 PI 控制方法，所提出的 MPC 控制器的频率波动更小，收敛速度更快。尤其是在 90s 时刻，系统出现大扰动，三种控制器的最大频率偏差分别是 $1.926\times10^{-3}$ p.u.、$1.42\times10^{-3}$ p.u.和 $2.453\times10^{-4}$ p.u.，标准差为 $5.194\times10^{-3}$、$3.595\times10^{-3}$、$3.788\times10^{-4}$。

## 4.2 风速、光照和负荷波动场景

实际工作中，风速、光照和系统负荷处在波动状态。本场景将 3 分钟实际风速、光照和负荷波动数据导入仿真系统。风、光伏波动和负荷波动如下图 6a 所示。该情景下的仿真结果见图 6b~图 6f，其中实线代表实际功率波动，虚线代表对于扰动的观测辨识；黑色表示负荷扰动，蓝色表示光伏功率波动，橙色表示风机功率波动。

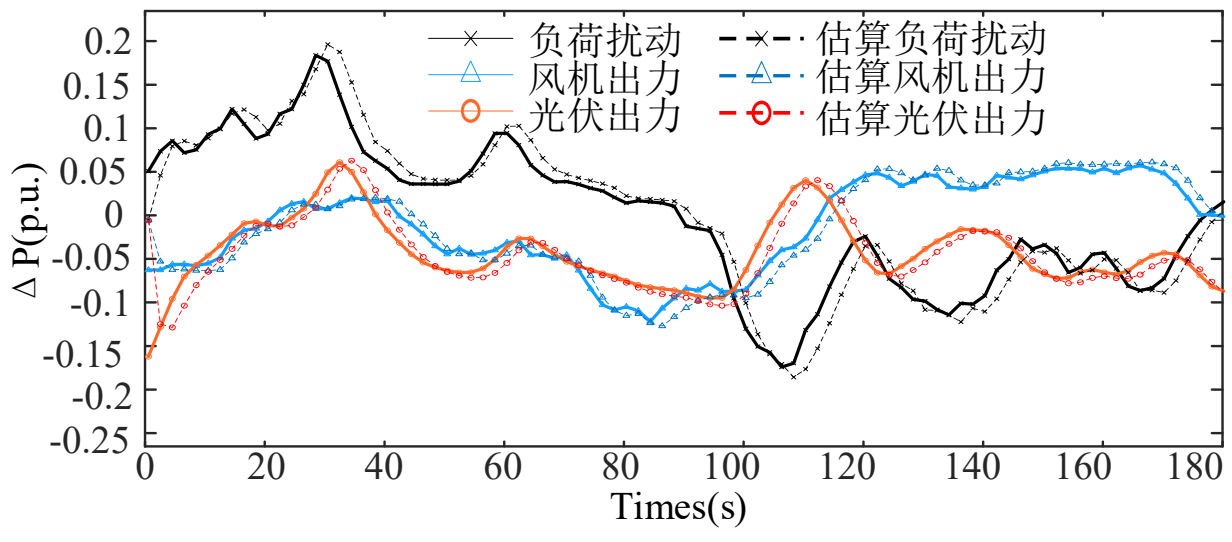

a.功率波动

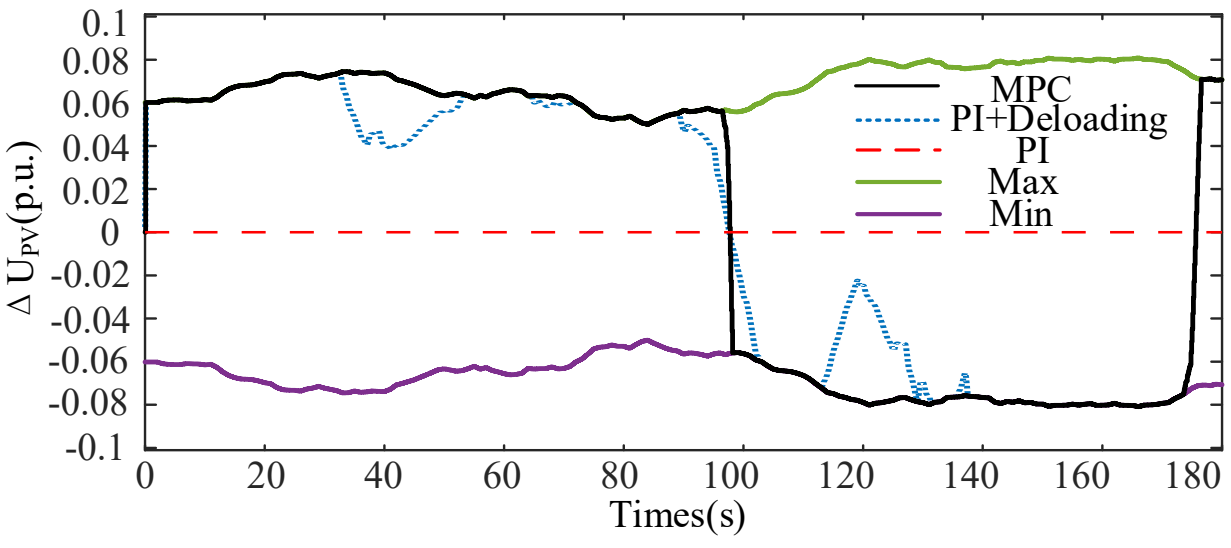

b.光伏出力

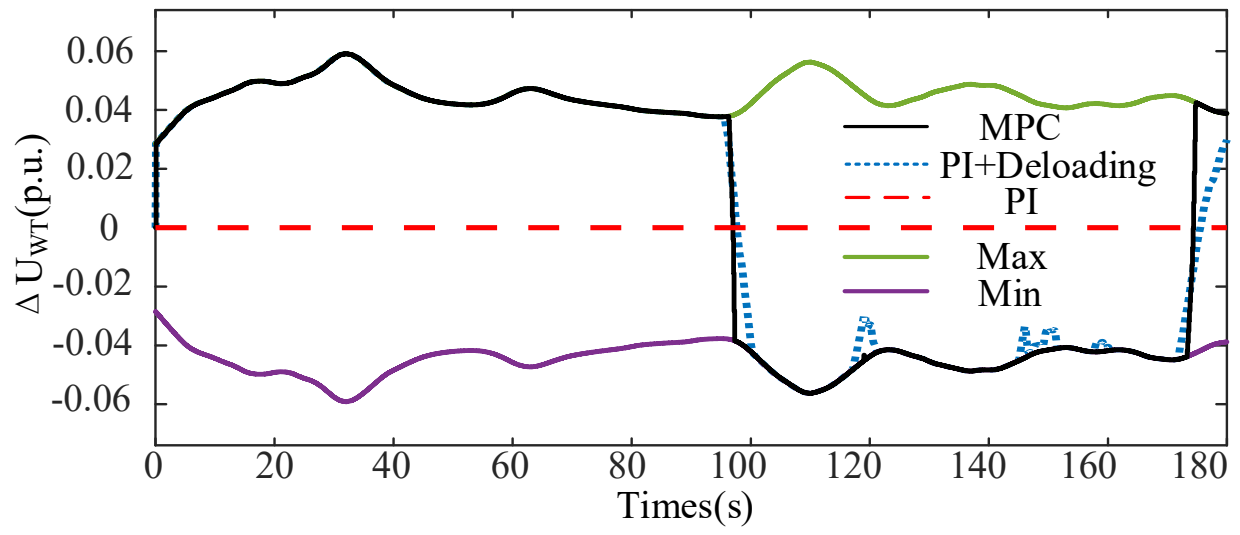

c.风机出力

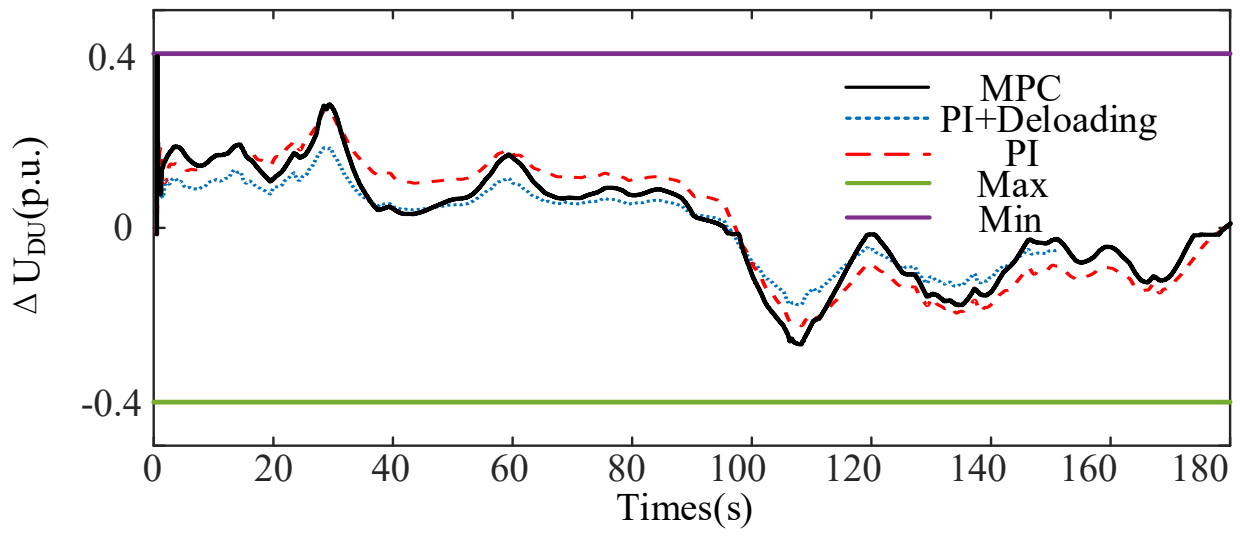

d.柴油机出力

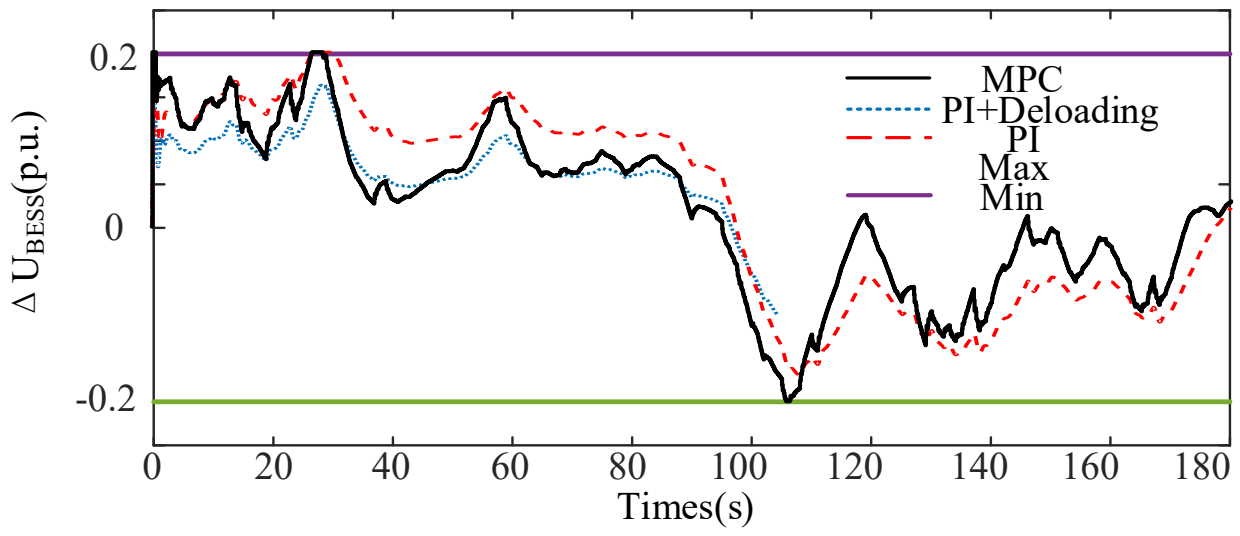

e.储能系统出力

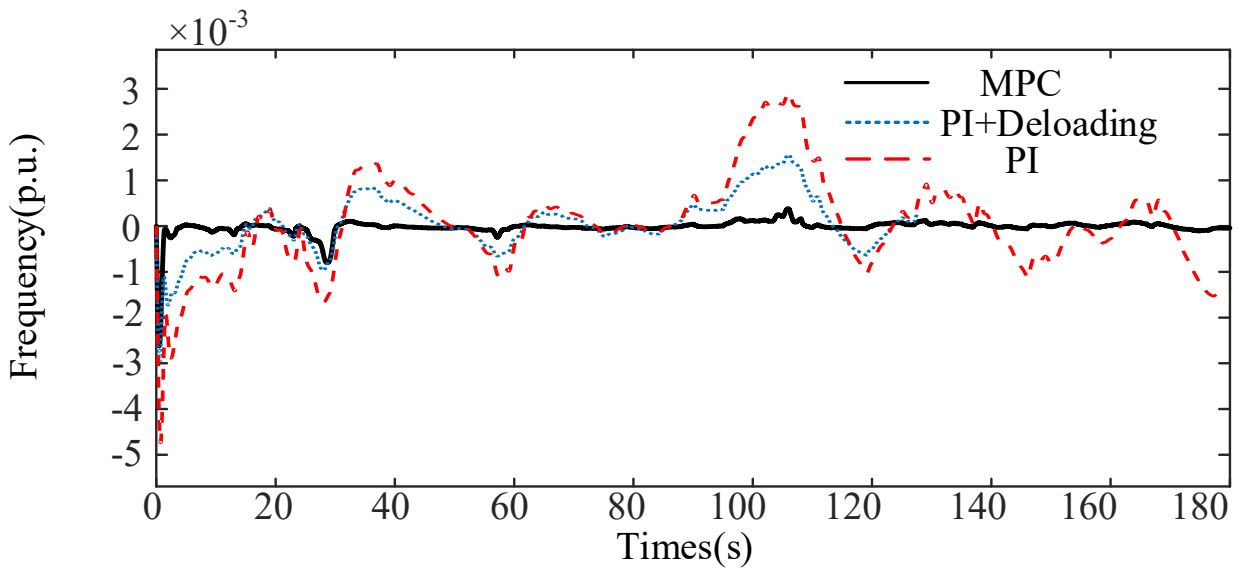

f.系统频率

图 6 案例 2 的结果

Fig.6 Case 2 Results

受风、光伏波动性的影响，其可用调频功率限值也出现波动，见图 6b~图 6c。两种 PI 控制方法中，各个发电单元的出力形状一致，见图 6b~图 6e。而 MPC 控制，由于风、光惩罚因子更小，优先风、光发电单元参与调频。图 6a 中，当考虑风、光波动较平稳时，观测器所预测的值与实际值基本一致。图 6b~图 6c 的部分时间段，风、光调频出力达到限值。相应的，柴油机、储能调频出力得到减少，提高微电网经济运行。

三种控制方法中，MPC 二次调频控制方法最

优，如图 8f 所示。三种方法最大频率偏差分别是 2.885×10-3 p.u.、1.541×10-3 p.u.和 7.732×10-4 p.u.，频率波动标准差分别是 2.477×10-3、7.392×10-3 和 1.385×10-2。

### 4.3 风速、光照快速变化场景

某些恶劣天气下，风速、光照可能会处于快速变化中。例如云层的快速遮蔽，风速的突然变化等。负荷波动与场景 2 保持一致，但风速，光照选取实测数据中快速变化的部分。本场景的风速、光照和负荷波动曲线如下图 7a 所示，其中实线代表实际功率波动，虚线代表对于扰动的观测辨识；黑色表示负荷扰动，蓝色表示光伏功率波动，橙色表示风机功率波动。

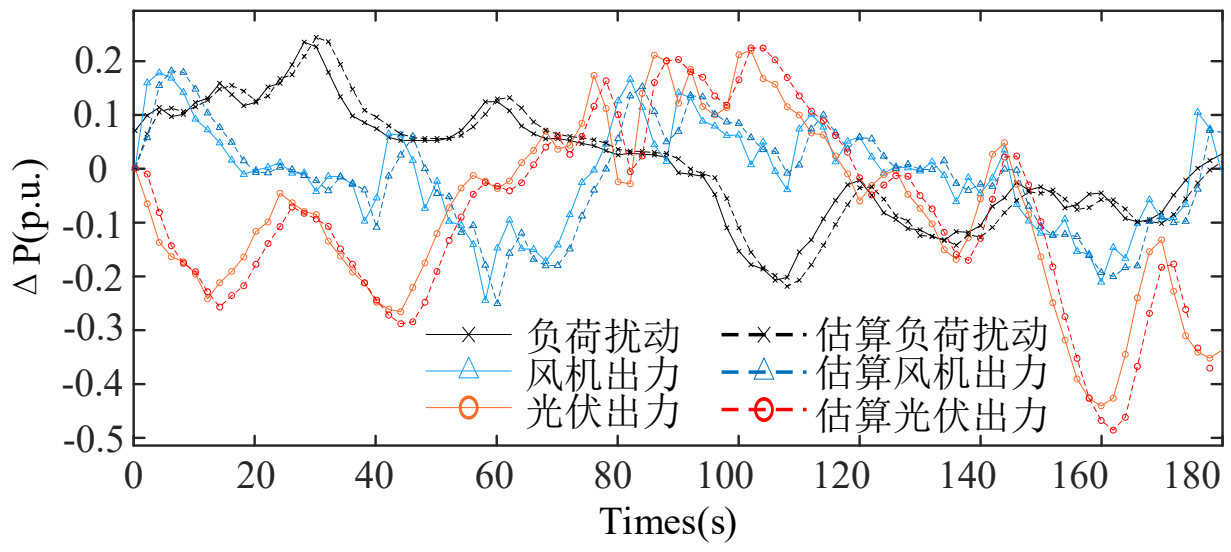

a.功率波动

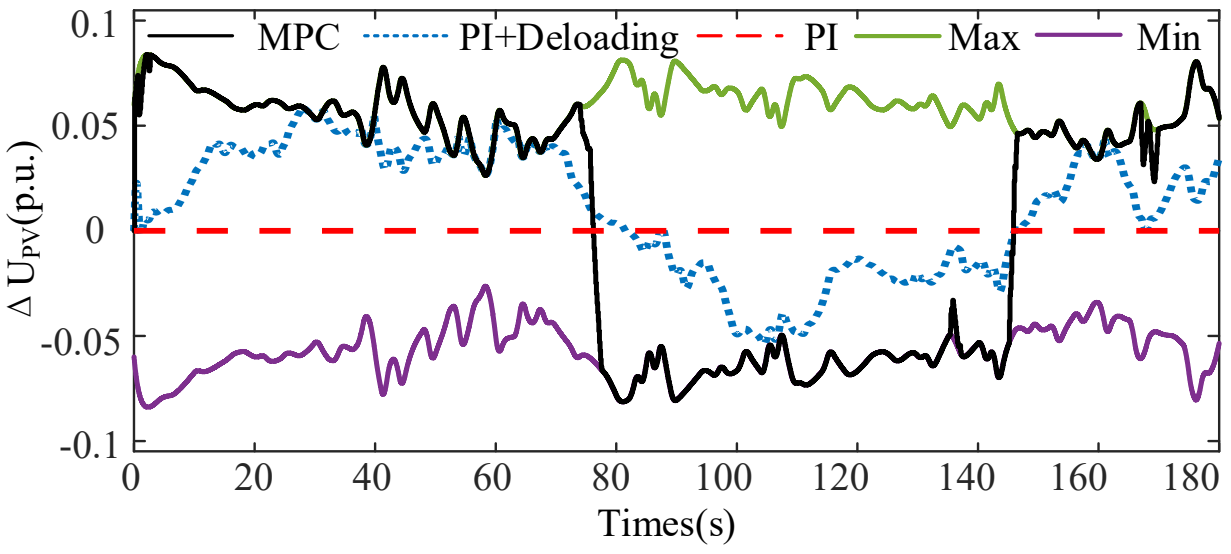

b.光伏出力

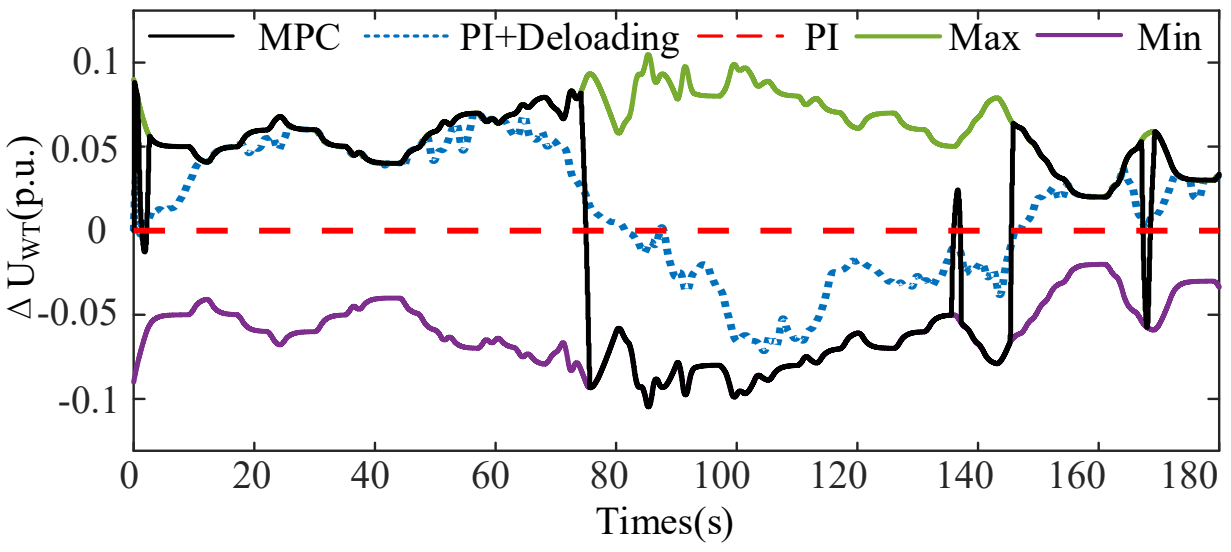

c.风机出力

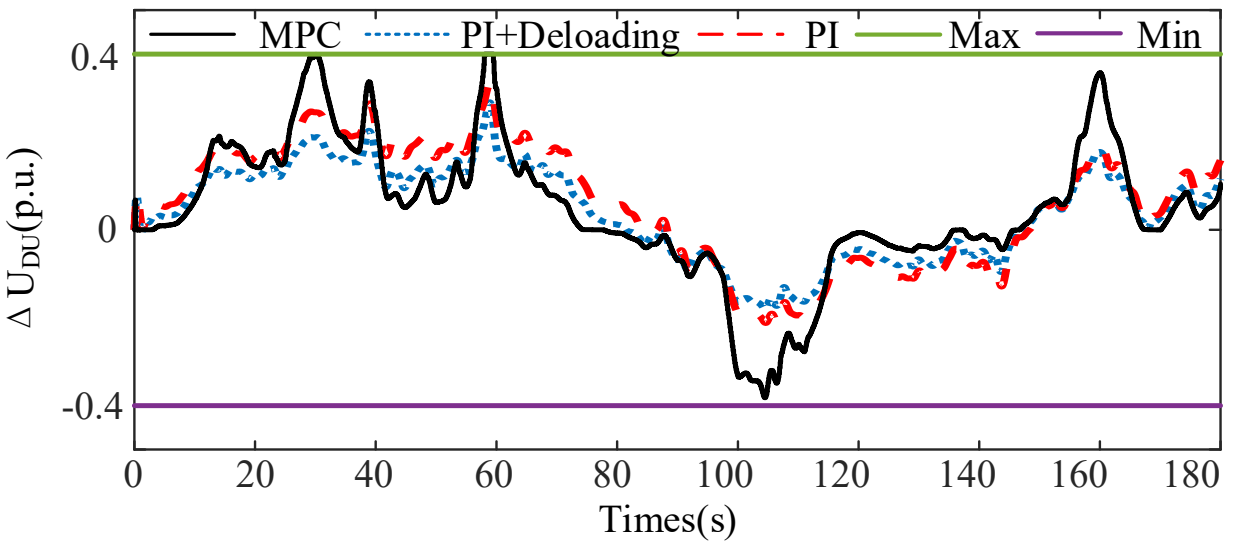

d.柴油机出力

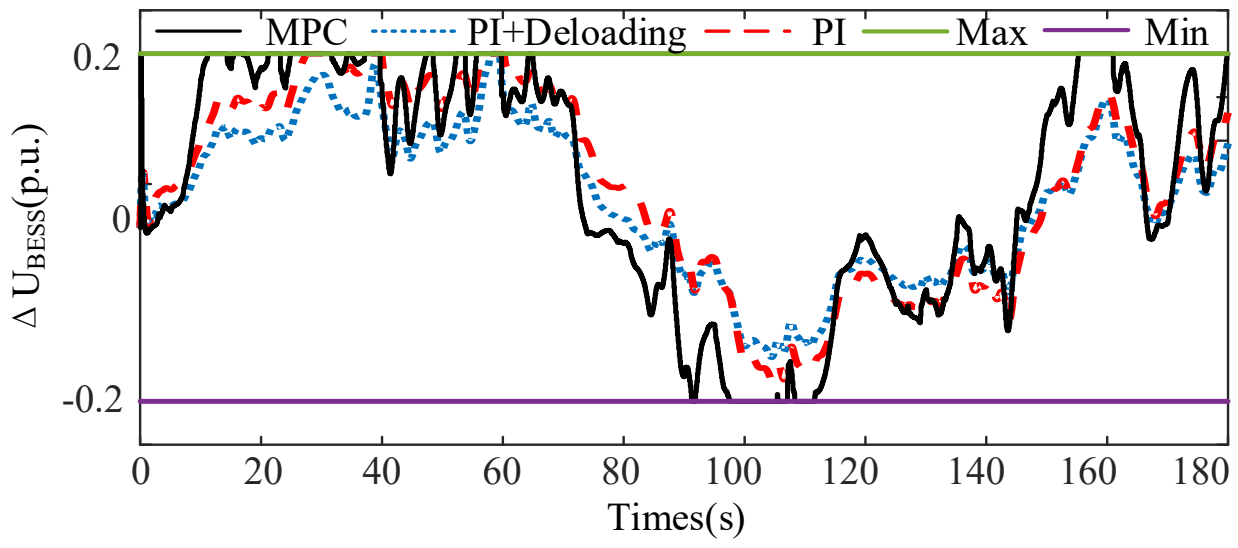

e.储能系统

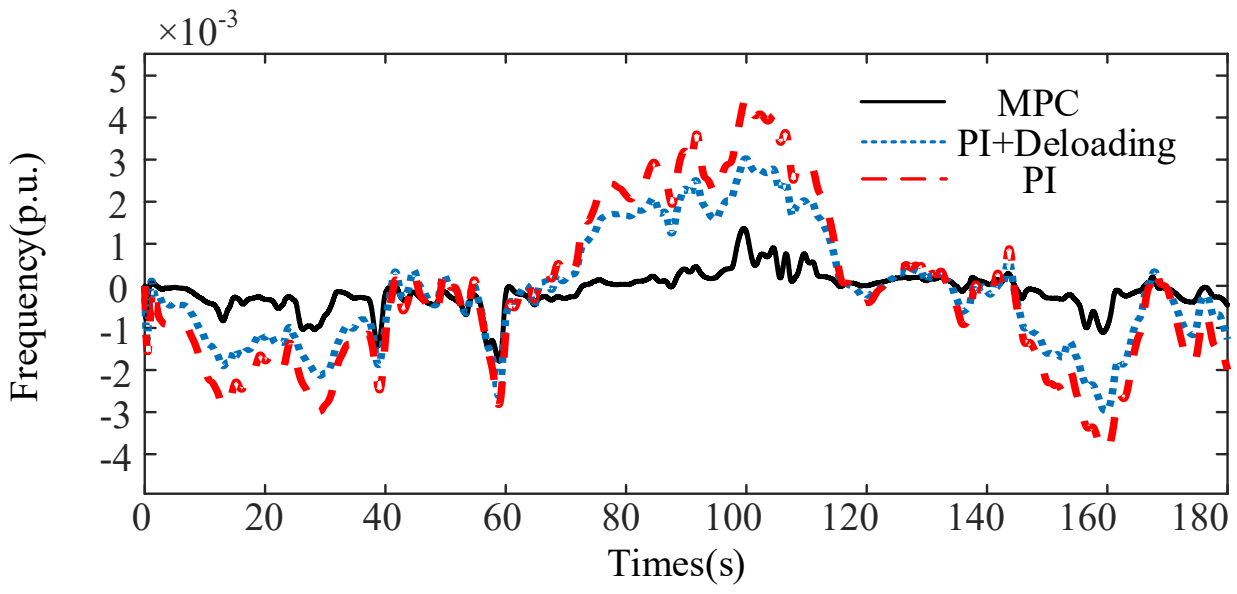

f.系统频率

图 7 案例 3 的结果

Fig.7 Case 3 Results

受光照、风速快速变化的影响。风电，光伏单元的可用调频功率存在剧烈波动动作。常规 PI 控制只依照机组额定容量，分配调频功率。PI+减载控制，部分区域的风、光机组的功率达到限值。由于饱和限值消减 PI 控制器部分输出，导致这些区域段的频率波动增大。

**表 3　三种控制器控制效果**

Tab.3 Three controller control effects

| 不同工况 | 控制方式 | 标准差 | 最大频率偏差 | 单位 |
|---|---|---|---|---|
| 案例 1 | MPC | $3.788\times10^{-4}$ | $2.453\times10^{-4}$ | p.u. |
| | PI+deloading | $3.595\times10^{-3}$ | $1.42\times10^{-3}$ | p.u. |
| | PI | $5.394\times10^{-3}$ | $1.92\times10^{-3}$ | p.u. |
| 案例 2 | MPC | $2.477\times10^{-3}$ | $7.732\times10^{-4}$ | p.u. |
| | PI+deloading | $7.392\times10^{-3}$ | $1.541\times10^{-3}$ | p.u. |
| | PI | $1.385\times10^{-2}$ | $2.88\times10^{-3}$ | p.u. |
| 案例 3 | MPC | $5.934\times10^{-3}$ | $1.584\times10^{-3}$ | p.u. |
| | PI+deloading | $1.818\times10^{-2}$ | $2.970\times10^{-3}$ | p.u. |
| | PI | $2.6\times10^{-2}$ | $4.563\times10^{-3}$ | p.u. |

相比较，提出的 MPC 控制方法，由于优先光伏、风电出力，部分区域风、光发电单元调频出力也达到上限值。但是，MPC 控制是在考虑功率约束下的最优控制输出。因此，这些区域段，柴油、储能输出功率会适当提高，来保证频率调节性能。因此，相比于场景 2，场景 3 中 MPC 二次调频方法的频率响应得到更明显程提升。三种控

制方法的最大频率偏差分别为 $4.563\times10^{-3}$ p.u.、$2.970\times10^{-3}$ p.u.和 $1.584\times10^{-3}$ p.u.，标准差分别为 $2.6\times10^{-2}$、$1.818\times10^{-2}$ 和 $5.934\times10^{-3}$。

为了更加直观的体现所提出控制器在二次调频控制响应性能，表 3 描述了三种控制器的最大频率偏差以及标准差。

## 5 结论

微电网二次调频控制是个多输入单输出且输入存在随机性的系统控制问题。针对随机性的问题，引入不确定输入扰动观测，对未知负荷和发电扰动进行卡尔曼滤波观测；增加风、光伏发电实时约束，避免机组调频超出功率限额；通过权重系数设置，优先风、光发电参与二次调频。所提 MPC 控制器在线优化调整各个分布式电源的出力，提供微电网的频率控制能力。建立了含多台风、光发电的独立微电网二次调频模型。风、光伏恒定，典型风、光伏波动和剧烈风、光波动场景下，对所提方法进行了对比验证。结果显示所提方法能够加快系统的频率恢复速度，减小频率的波动峰值，尤其在剧烈风、光波动场景下，频率调节提升效果更为明显。

## 参考文献

# SECONDARY FREQUENCY CONTROL OF ISLANDED MICROGRID CONSIDERING WIND AND SOLAR STOCHASTICS

Zhong Cheng[1,2], Jiang Zhifu[2], Zhang Xiangyu[2], Chen Jikai[1,2], Li Yang[1,2]

(1. *Key Laboratory of Modern Power System Simulation and Control &Renewable Energy Technology, Ministry of Education(Northeast Electric Power University* Jilin 132012;

2.*Northeast Electric Power University, College of Electrical Engineering,* Jilin 132012)

**Abstract:** This paper proposed a model predictive control (MPC) secondary frequency control method considering wind and solar power generation stochastics. The extended state-space matrix including unknown stochastic power disturbance is established, and a Kalman filter is used to observe the unknown disturbance. The maximum available power of wind and solar DGs is estimated for establishing real- time variable constraints that prevent DGs output power from exceeding the limits. Through setting proper weight coefficients，wind and photovoltaic DGs are given priority to participate in secondary frequency control. The distributed restorative power of each DG is obtained by solving the quadratic programming (QP) optimal problem with variable constraints. Finally，a microgrid simulation model including multiple PV and wind DGs is built and performed in various scenarios compared to the traditional secondary frequency control method. The simulation results validated that the proposed method can enhance the frequency recovery speed and reduce the frequency deviation，especially in severe photovoltaic and wind fluctuations scenarios.

**Keywords:** Secondary Frequency Control; model predictive control; islanded microgrid; deloading control; stoch astic input observer

# 附录

$$A_{\mathrm{c}}=\begin{bmatrix}
0 & 0 & \frac{1}{2H_{\mathrm{t}}} & 0 & \frac{1}{2H_{\mathrm{t}}} & \frac{1}{2H_{\mathrm{t}}} & \frac{1}{2H_{\mathrm{t}}} & 0 & \frac{1}{2H_{\mathrm{t}}} & \frac{1}{2H_{\mathrm{t}}} \\
0 & -\frac{1}{T_{\mathrm{pv1}}} & 0 & 0 & 0 & 0 & 0 & 0 & 0 & 0 \\
0 & \frac{1}{T_{\mathrm{pv2}}} & -\frac{1}{T_{\mathrm{pv2}}} & 0 & 0 & 0 & 0 & 0 & 0 & 0 \\
0 & 0 & 0 & -\frac{1}{T_{\mathrm{pv1}}} & 0 & 0 & 0 & 0 & 0 & 0 \\
0 & 0 & 0 & \frac{1}{T_{\mathrm{pv2}}} & -\frac{1}{T_{\mathrm{pv2}}} & 0 & 0 & 0 & 0 & 0 \\
0 & 0 & 0 & 0 & 0 & -\frac{1}{T_{\mathrm{wt}}} & 0 & 0 & 0 & 0 \\
0 & 0 & 0 & 0 & 0 & 0 & -\frac{1}{T_{\mathrm{wt}}} & 0 & 0 & 0 \\
-\frac{1}{RT_{\mathrm{DU1}}} & 0 & 0 & 0 & 0 & 0 & 0 & -\frac{1}{T_{\mathrm{DU1}}} & 0 & 0 \\
0 & 0 & 0 & 0 & 0 & 0 & -\frac{1}{T_{\mathrm{DU2}}} & -\frac{1}{T_{\mathrm{DU2}}} & 0 & 0 \\
-1 & 0 & 0 & 0 & 0 & 0 & 0 & 0 & 0 & -\frac{1}{T_{\mathrm{BESS}}}
\end{bmatrix}$$

$$B_{\mathrm{cu}}=\begin{bmatrix}
0 & 0 & 0 & 0 & 0 & 0 \\
\frac{1}{T_{\mathrm{pv1}}} & 0 & 0 & 0 & 0 & 0 \\
0 & 0 & 0 & 0 & 0 & 0 \\
0 & \frac{1}{T_{\mathrm{pv1}}} & 0 & 0 & 0 & 0 \\
0 & 0 & 0 & 0 & 0 & 0 \\
0 & 0 & \frac{1}{T_{\mathrm{wt}}} & 0 & 0 & 0 \\
0 & 0 & 0 & \frac{1}{T_{\mathrm{wt}}} & 0 & 0 \\
0 & 0 & 0 & \frac{1}{T_{\mathrm{DU1}}} & 0 & 0 \\
0 & 0 & 0 & 0 & 0 & 0 \\
0 & 0 & 0 & 0 & 0 & \frac{1}{T_{\mathrm{BESS}}}
\end{bmatrix}$$

$$C_{\mathrm{c}}=\begin{bmatrix}1 & 0 & 0 & 0 & 0 & 0 & 0 & 0 & 0 & 0\end{bmatrix}^{\mathrm{T}}$$

$$D_{\mathrm{cd}}=\begin{bmatrix}-\frac{1}{2H_{\mathrm{t}}} & -\frac{1}{2H_{\mathrm{t}}} & -\frac{1}{2H_{\mathrm{t}}} & -\frac{1}{2H_{\mathrm{t}}} & -\frac{1}{2H_{\mathrm{t}}}\end{bmatrix}$$

$$x=\begin{bmatrix}\Delta f & \Delta X_1 & \Delta P_{\mathrm{pv1}} & \Delta X_2 & \Delta P_{\mathrm{pv2}} & \Delta P_{\mathrm{wt1}} & \Delta P_{\mathrm{wt2}} & \Delta X_3 & \Delta P_{\mathrm{DU}} & \Delta P_{\mathrm{BESS}}\end{bmatrix}^{\mathrm{T}}$$

$$u=\begin{bmatrix}\Delta u_{\mathrm{pv1}} & \Delta u_{\mathrm{pv2}} & \Delta u_{\mathrm{wt1}} & \Delta u_{\mathrm{wt2}} & \Delta u_{\mathrm{DU}} & \Delta u_{\mathrm{BESS}}\end{bmatrix}^{\mathrm{T}}$$

$$d=\begin{bmatrix}\Delta P_{\mathrm{D}} & \Delta P_{\mathrm{pv1}} & \Delta P_{\mathrm{pv2}} & \Delta P_{\mathrm{wt1}} & \Delta P_{\mathrm{wt2}}\end{bmatrix}^{\mathrm{T}}$$

$$y=\Delta f$$

式中:$\Delta f$ 为频率偏差，$\Delta x_1$ 为光伏阵列输出功率，$\Delta P\mathrm{pv}$ 为光伏逆变器输出功率，$\Delta P\mathrm{wt}$ 为风机输出功率，$\Delta x_2$ 为调速器增量，$\Delta P_{\mathrm{DU}}$ 为柴油机输出功率，$\Delta P_{\mathrm{BESS}}$ 为储能系统输出功率，$H_{\mathrm{t}}$为等效惯性。